# Scale-free Distributed Cooperative Voltage Control of Inverter-based Microgrids with General Time-varying Communication Graphs

Donya Nojavanzadeh, *Student Member, IEEE*, Saeed Lotfifard, *Senior Member, IEEE*, Zhenwei Liu, *Member, IEEE*, Ali Saberi, *Life Fellow, IEEE*, and Anton A. Stoorvogel, *Senior Member, IEEE*

*Abstract*— **This paper presents a method for controlling the voltage of inverter-based Microgrids by proposing a new scale-free distributed cooperative controller. The communication network is modeled by a general time-varying graph which enhances the resilience of the proposed protocol against communication link failure, data packet loss, and fast plug and play operation in the presence of arbitrarily communication delays. The proposed scale-free distributed cooperative controller is independent of any information about the communication system and the size of the network (i.e., the number of distributed generators). The stability analysis of the proposed protocol is provided. The proposed method is simulated on the CIGRE medium voltage Microgrid test system. The simulation results demonstrate the feasibility of the proposed scale-free distributed nonlinear protocol for regulating voltage of Microgrids in presence of communication failures, data packet loss, noise, and degradation.**

## I. INTRODUCTION

TO create a reliable Microgrid a variety of issues should be addressed such as seamless transition from the grid connected mode to the islanded mode [1], frequency control [2] and energy management of the microgrids [3]. Voltage control is also a challenging task for realizing reliable Microgrids, which is the focus of this paper. The voltage of a pilot node/bus or the average value of voltages of multiple pilot buses is regulated to a reference voltage value. Droop-based control is the most common method for controlling the voltage of Microgrids in which each distributed generator contributes to the voltage control according to its droop gain value. However, droop controllers cannot eliminate the steady state error of the regulated voltage signal. Therefore, a secondary controller is utilized to eliminate this steady state error. The conventional secondary controllers have a centralized architecture in which a single centralized controller generates the control signal and sends the control commands to each individual distributed generator. This scheme suffers from the single-point-of-failure problem and requires a fast communication system between every distributed generator and the centralized controller.

D. Nojavanzadeh, S. Lotfifard and A. Saberi are with the school of Electrical Engineering and Computer Science, Washington State University, Pullman, WA 99164, USA

Z. Liu is with the College of Information Science and Engineering, Northeastern University, Shenyang 110819, P. R. China.

A. A. Stoorvogel is with the Department of Electrical Engineering, Mathematics and Computer Science, University of Twente, P.O. Box 217, Enschede, The Netherlands.

In recent years, distributed cooperative controllers have been utilized for power systems applications where they only require peer-to-peer communications among agents (i.e., distributed generators). The single- or double-integrator agent models for distributed generators are derived and used in the distributed cooperative controllers. Note that in the case of modeling the agents by single-integrator dynamics, the static gain is a well-known and classical controller vastly utilized as the secondary controller for Microgrids [4-7]. A distributed controller for Microgrid with fixed communication graph is utilized in [5] that is based on model predictive control. In [6] a distributed-averaging proportional integral (DAPI) controller is developed for control of islanded Microgrids. Secondary control of Microgrids in the presence of uncertainties in communication links is studied in [7] where the discrete-time protocols depend on some knowledge of the communication network. It should be mentioned that in [7], authors assumed a dwell-time for the switching communication topology. This means if the communication topology changes faster than the assumed dwell-time the stability of the distributed controller is endangered. Distributed secondary control of Microgrids in the presence of communication delays is considered in [8], where individual delay bounds with respect to different sets of controller gains and pinning is required. It is worth to note that single-integrator dynamic is neutrally stable, therefore the proposed controllers in the literature for systems modeled by single-integrator agents are scale-free and independent from the knowledge of the communication network, however, the associated communication graph for these controllers is required to be *fixed* meaning that topology of the communication network should remain fixed for sufficiently large duration. It is well-known that protocols which is designed for fixed graphs does not guarantee stability or consensus for dealing with the cases that switching occurs arbitrarily in the network. In the case of switching graphs, usually one should know the knowledge about dwelling-time to design protocols. To address the abovementioned problem, a nonlinear protocol is proposed in [9] for distributed cooperative control of multi-terminal HVDC systems with time-varying communication networks.

On the other hand, for Microgrid systems with double-integrator agents, unlike the single-integrator case, the proposed linear controllers in the literature require some knowledge about the associated communication graph such as lower bound on the spectrum of the associated Laplacian matrix and the size of the network [10-12], as such unlike the case of single-integrator, the available linear protocols for networks with double-integrators agents the stability is not quaranteed for cases that any switching occur in the network. Therefore, some nonlinear protocols based on adaptive methodology have been proposed to obviate the requirement for the knowledge of the communication graphs [13-14].

One of the challenges for the practical implementation of distributed cooperative controllers is that the structure of communication systems is variable due to a variety of reasons such as possible outages of communication links, data packet loss or noise in the communication systems. Therefore, proposing a scale-free distributed cooperative controller for networks with time-varying communication graph and independent from the knowledge of the communication network is of special interest.

In this paper, a distributed scale-free nonlinear protocol for voltage control of Microgrids is proposed with the following main contributions: (1) The proposed secondary controller is designed for Microgrid systems with time-varying communication topologies which includes switching networks as well. This feature enhances the resilience of the protocol in practical application in which the networks are subject to communication link failures, data packet drops and noisy communication networks. The classical techniques for dealing with the switched systems such as high-gain protocols [15] require considering the dwell-time for the stability. For instance, data packet drops may occur with any rate. One of the strengths of the proposed protocol in this paper is independence from the communication network and the dwell-time. (2) The proposed distributed controller is developed for the networks subject to unknown, non-uniform and arbitrarily communication delays where the stability of the system is analyzed mathematically. (3) The designed scale-free nonlinear dynamic controller does not require any knowledge of the communication structure such as the knowledge of the second smallest real part of eigenvalues of the Laplacian matrix associated with the communication graph and the size of the network (i.e., the number of distributed generators). Therefore, the controller can operate properly despite the changes in the communication architecture. This feature improves the robustness of the controller against possible communication failures that may lead to the change in the communication network structure and the change in the number of distributed generators controlled by the distributed cooperative controller.

The rest of this paper is organized as follows. The model of IIDGs for the distributed controller is presented in section II. The proposed distributed cooperative controller for voltage control of Microgrids is explained in section III. In section IV, simulation results are presented and finally section V concludes the paper.

## II. Modelling of Inverter Interfaced Distributed Generator (IIDG) for Distributed Control

Fig.1 shows the typical controllers of inverter interfaced distributed generators. The fundamental component of the output reactive power of an IIDG is represented as follows [16]

$$\dot{Q}^i = -\omega_c Q^i + \omega_c \tilde{Q}^i \quad (1)$$

where $\omega_c$ is the cutoff frequency of the low-pass filter [16] and

$$\tilde{Q}^i = V_{oq}^i I_{od}^i - V_{od}^i I_{oq}^i \quad (2)$$

where $V_{oq}^i$ and $V_{od}^i$ denote the quadrature-axis and direct-axis components of the output voltage of i[th] IIDG, respectively. $I_{oq}^i$ and $I_{od}^i$ represent the quadrature-axis and direct-axis components of the output current of i[th] IIDG, respectively.

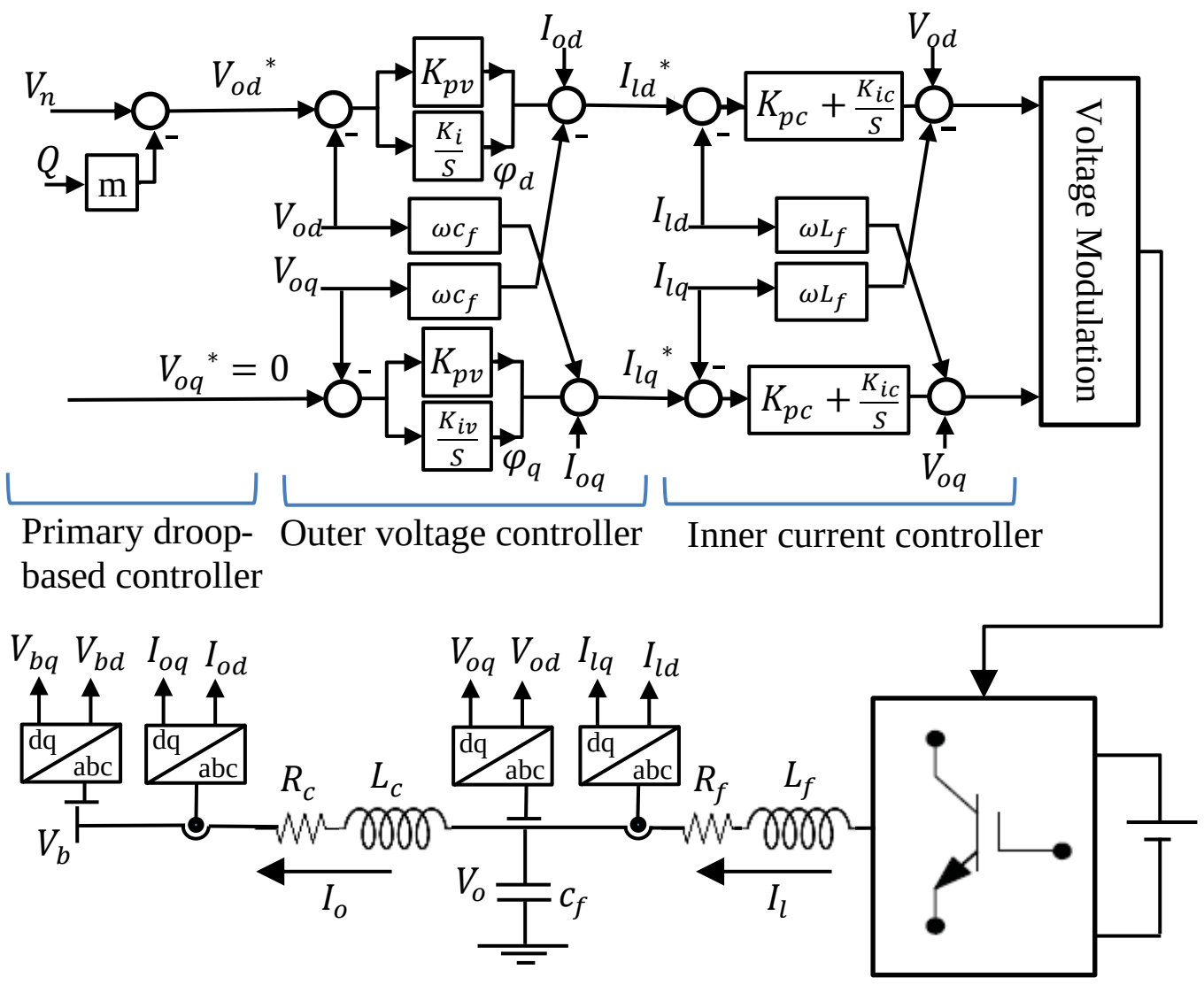

Fig.1. Controllers of inverter interfaced distributed generators

From (1), also it follows

$$\ddot{Q}^i = -\omega_c \dot{Q}^i + \omega_c \dot{\tilde{Q}}^i \quad (3)$$

where

$$\dot{\tilde{Q}}^i = \dot{V}_{oq}^i I_{od}^i + V_{oq}^i \dot{I}_{od}^i - \dot{V}_{od}^i I_{oq}^i - V_{od}^i \dot{I}_{oq}^i \quad (4)$$

where

$$\dot{V}_{od}^i = \omega V_{oq}^i + \frac{1}{c_f} I_{ld}^i - \frac{1}{c_f} I_{od}^i \quad (5)$$

$$\dot{V}_{oq}^i = -\omega V_{od}^i + \frac{1}{c_f} I_{lq}^i - \frac{1}{c_f} I_{oq}^i \quad (6)$$

$$\dot{I}_{od}^i = \frac{-r_c}{L_f} I_{od}^i + \omega I_{oq}^i + \frac{1}{L_c} V_{od}^i - \frac{1}{L_c} V_{bd}^i \quad (7)$$

$$\dot{I}_{oq}^i = \frac{-r_c}{L_f} I_{oq}^i - \omega I_{od}^i + \frac{1}{L_c} V_{oq}^i - \frac{1}{L_c} V_{bq}^i \quad (8)$$

where $r_c$ represents the coupling resistance of the IIDG and $r_f$, $L_f$, $c_f$ denote the resistance, inductance and capacitance of the output filter of IIDG, respectively.

The inner current loop of inverters has a very fast dynamic response. Therefore, (5) and (6) can be rewritten as follows

$$\dot{V}_{od}^i = \omega V_{oq}^i + \frac{1}{c_f} {I_{ld}^*}^i - \frac{1}{c_f} I_{od}^i \quad (9)$$

$$\dot{V}_{oq}^i = -\omega V_{od}^i + \frac{1}{c_f} {I_{lq}^*}^i - \frac{1}{c_f} I_{oq}^i \quad (10)$$

where

$${I_{ld}^*}^i = I_{od}^i - \omega c_f V_{oq}^i + K_{pv}^i\left({V_{od}^*}^i - V_{od}^i\right) + \varphi_d^i \quad (11)$$

$${I_{lq}^*}^i = I_{oq}^i + \omega c_f V_{od}^i + K_{pv}^i\left({V_{oq}^*}^i - V_{oq}^i\right) + \varphi_q^i \quad (12)$$

where

$${V_{od}^*}^i = V_n^i - m^i Q^i \quad (13)$$

$${V_{oq}^*}^i = 0 \quad (14)$$

where $m^i$ denotes the droop gain of i^th IIDG, $K_{pc}$ and $K_{ic}$ represent the proportional and integral gains of the inner current controller of IIDG, respectively. $K_{pv}$ and $K_{iv}$ denote the proportional and integral gains of the outer voltage controller of IIDG, respectively. $V_n^i$ represents the nominal set point of direct-axis component of the output voltage of i^th IIDG. ${V_{oq}^*}^i$ and ${V_{od}^*}^i$ represent the set-points of the quadrature-axis and direct-axis components of the output voltage of i^th IIDG, respectively. ${I_{od}^*}^i$ and ${I_{oq}^*}^i$ represent the set-points of direct-axis and quadrature-axis components of the output current of i^th IIDG, respectively. $I_{ld}^i$and $I_{lq}^i$ represent the direct-axis and quadrature-axis components of the current flowing through the output filter of i^th IIDG, respectively. ${I_{ld}^*}^i$ and ${I_{lq}^*}^i$ denote the set-point of direct-axis and quadrature-axis components of the current flowing through the output filter of i^th IIDG.

By replacing (1), (4)-(14) into (3), the following holds

$$\ddot{Q}^i = u'^i \tag{15}$$

where $u'^i$ is function of $V_n^i$ and locally available measurements. Therefore, once the distributed cooperative controller determines the value of $u'^i$, the value of $V_n^i$ can be calculated accordingly.

Equation (15) can be rewritten as

$$\Delta\ddot{Q}^i = u'^i \tag{16}$$

where $\Delta Q^i$=$Q^i - Q_0^i$ and $Q_0^i$ is the latest dispatched set-point of the i^th IIDG.

The main goal of voltage control of Microgrids is to regulate the voltage of a pilot bus or the average value of voltages of multiple pilot buses while available IIDGs are utilized according to the defined participation factors. In defining the participation factor of each IIDG, different aspects such as the location of the IIDGs and their capacities should be taken into account. Assuming the participation factor of i^th IIDG is $\alpha^i$, the following relationship should hold

$$\frac{\Delta Q^1}{\alpha^1} = \frac{\Delta Q^2}{\alpha^2} = \cdots = \frac{\Delta Q^i}{\alpha^i} = \Delta Q_{ref} \tag{17}$$

for $i = 1, \dots, N$. This means to regulate the voltage signal, the change in the output reactive power of an IIDG with a bigger participation factor should be larger than the change in the output reactive power of an IIDG with a smaller participation factor.

According to the above discussion, the voltage control objective is defined as follows

$$\lim_{t\to\infty}\left(\frac{\Delta Q^i}{\alpha^i} - \Delta Q_{ref}\right) = 0, \tag{18}$$

for $i = 1, \dots, N$, where $\Delta Q_{ref}$ is calculated at the lead IIDG as follows

$$\Delta Q_{ref} = K_p\left(V_{pilot} - V_{ref}\right) + K_i \int \left(V_{pilot} - V_{ref}\right) \tag{19}$$

Then, equation (16) can be rewritten as

$$\dot{x}^i = Ax^i + Bu^i \tag{20}$$

where $A = \begin{pmatrix} 0 & 1 \\ 0 & 0 \end{pmatrix}, B = \begin{pmatrix} 0 \\ 1 \end{pmatrix}$ an $x^i = \begin{bmatrix} \frac{\Delta Q^i}{\alpha^i} & \frac{\Delta\dot{Q}^i}{\alpha^i} \end{bmatrix}^T$ and $u^i = \frac{u'^i}{\alpha^i}$ for $i = 1, \dots, N$.

## III. Distributed Cooperative Control of Inverter Interfaced Distributed Generators (IIDGs)

Consider a Microgrid system with $N$ IIDGs modeled by (20), communicating through a time-varying communication graph. Each IIDG $i = 1, \dots, N$, has access to the following information which is a linear combination of its own state relative to that of neighboring IIDGs where the communication network is subject to delays.

$$\zeta^i(t) = \sum_{j=1}^{N} a_{ij}(t)\left(x^i(t) - x^j\left(t - \tau_{ij}\right)\right) \tag{21}$$

where $a_{ij}(t) \geq 0$, $a_{ij}(t) = a_{ji}(t)$ and $a_{ii}(t) = 0$, for all $t$, and $\tau_{ij} \in \mathbb{R}_{\geq 0}$ represents an unknown communication delay from agent $j$ to agent $i$ and $\tau_{ii}$=0.

The Laplacian matrix associated to the communication graph is defined as

$$\ell_{ij}(t) = \begin{cases} \sum_{k=1}^{N} a_{ik}(t), & i = j, \\ -a_{ij}(t), & i \neq j \end{cases} \tag{22}$$

The review of graph theory can be found in [17,18]. In terms of Laplacian matrix, (21) can be rewritten as

$$\zeta^i(t) = \sum_{j=1}^{N} \ell_{ij}(t) x^j\left(t - \tau_{ij}\right). \tag{23}$$

which implies that all states are communicated among all IIDGs. The objective is that voltage of IIDGs regulated to a priori given reference trajectory $x^r = [\Delta Q_{ref} \quad 0]^T$. When IIDGs have access to the reference trajectory $x^r$ the available information can be rewritten as follows

$$\bar{\zeta}^i(t) = \zeta^i(t) + \left(x^i(t) - x^r(t)\right) \tag{24}$$

otherwise, the IIDG has the same information as before

$$\bar{\zeta}^i(t) = \zeta^i(t). \tag{25}$$

Therefore, the following can be written

$$\begin{aligned} &\bar{\zeta}^i(t) = \zeta^i(t) + \iota^i\left(x^i(t) - x^r(t)\right) \\ &\iota^i = \begin{cases} 1, & x^r \text{ available} \\ 0, & x^r \text{ unavailable} \end{cases} \qquad i = 1, \dots, N. \end{aligned} \tag{26}$$

The expanded Laplacian matrix is defined as follows:

$$\bar{L}(t) = L(t) + diag\{\iota^i(t)\} = \left[\bar{\ell}_{ij}\right]_{N\times N} \tag{27}$$

We make the following definition for time-varying graphs.

**Definition 1 [19]** For given real numbers $\gamma > \beta > 0$, $\mathbb{G}_{\beta,\gamma,N}$ denotes the set of all time-varying, weighted, and undirected graphs satisfying

$$\beta I \leq \bar{L}(t) \leq \gamma I \tag{28}$$

for all $t$.

**Proposed protocol:** A scale-free adaptive nonlinear protocol is proposed as follows for each IIDG $i = 1, \dots, N$

$$\begin{aligned} \dot{\rho}^i &= {\bar{\zeta}^i}^T PBB^TP\bar{\zeta}^i \\ u^i &= -\rho^i B^TP\bar{\zeta}^i \end{aligned} \tag{29}$$

where $\rho^i(0) \geq 0$ and $P > 0$ is the unique solution of

$$PA + A^TP - PBB^TP + M = 0, \tag{30}$$

where $M$ is any positive definite matrix.

In the following we provide two theorems related to the proposed protocol. The first theorem is for Microgrids with associated general time-varying communication graphs, and the second theorem is for the case of switched communication graphs.

**Theorem 1 (general time-varying graph)** Consider a Microgrid with IIDGs described by (20) and communication information (26) and the reference trajectory $x^r = [\Delta Q_{ref} \quad 0]^T$. Let $\mathbb{G}_{\beta,\gamma,N}$ denote a set of time-varying graph as defined in Definition 1. Then, the scale-free nonlinear protocol (29) achieves voltage regulation with IIDGs participation factors (18) in the presence of communication delays, for the reference trajectory $x^r$ and any graph $\mathcal{G}(t) \in \mathbb{G}_{\beta,\gamma,N}$ with any $\gamma > \beta > 0$, any size of the network and any communication delay $\tau_{ij} \in \mathbb{R}_{\geq 0}$.

*Proof:* The proof is given in Appendix A.

**Definition 2** *Let $\mathcal{L}_N \subset \mathbb{R}^{N\times N}$ be any family of finite number of Laplacian matrices associated to undirected, weighted and strongly connected graphs with N agents. Let $\mathcal{G}_L$ denote the graph associated with a Laplacian matrix $L \in \mathcal{L}_N$. Then, a time-varying graph $\mathcal{G}(t) \in \mathbb{G}_N$ with N agents has definition as*

$$\mathcal{G}(t) = \mathcal{G}_{\sigma(t)},$$

*where $\sigma: \mathbb{R} \to \mathcal{L}_N$ is a piecewise constant and right-continuous function with arbitrary small minimal dwell-time $\tau$ (see [20]) i.e., $\sigma(t)$ remains fixed for $t \in [t_k, t_{k+1}), k \in \mathbb{Z}$ and switches at $t = t_k$ , $k = 1,2,\ldots,$ where $t_{k+1} - t_k \geq \tau$ for $k = 0,1,\ldots.$ We assume $t_0 = 0$.*

**Theorem 2 (switched graph)** *For a Microgrid that includes IIDGs described by agent models (20), communication information (26), the reference trajectory $x^r = [\Delta Q_{ref} \quad 0]^T$ and the switching communication topology $\mathcal{G}_{\sigma(t)} \in \mathbb{G}_N$ as defined in Definition 2, the scale-free nonlinear protocol (29) achieves voltage regulation with participation factors (18) for any graph $\mathcal{G}_{\sigma(t)} \in \mathbb{G}_N$ with any possible $\mathcal{L}_N$, any size of the network, any communication delay $\tau_{ij} \in \mathbb{R}_{\geq 0}$, and any reference trajectory $x^r = (\Delta Q_{ref} \quad 0)^T$.*

*Proof:* The proof is given in Appendix B.

## IV. SIMULATION RESULTS

In this section, the feasibility and performance of the proposed distributed cooperative controller for voltage control of Microgrids is demonstrated. As shown in Fig. 2, the GIGRE medium voltage test system [21] is utilized and simulated in MATLAB/Simulink. Four 4.3 $MVA$, 630 $V$ IIDGs are added to the Microgrid through $630/1247\ V$ transformers. According to [22], IIDGs have $CF = 4700$ µF, $Lf = 0.023\ mH$, and $Rf = 0.849\ mOhms$ . Different scenarios are considered to demonstrate the effectiveness of the proposed method. As droop gains are commonly determined based on the capacity of the IIDGs, the participation factors are also determined based on the droop gains. This means in (18) $\alpha_i = \frac{1}{m_i}$ where $m_i$ is the droop gain of i[th] IIDG.

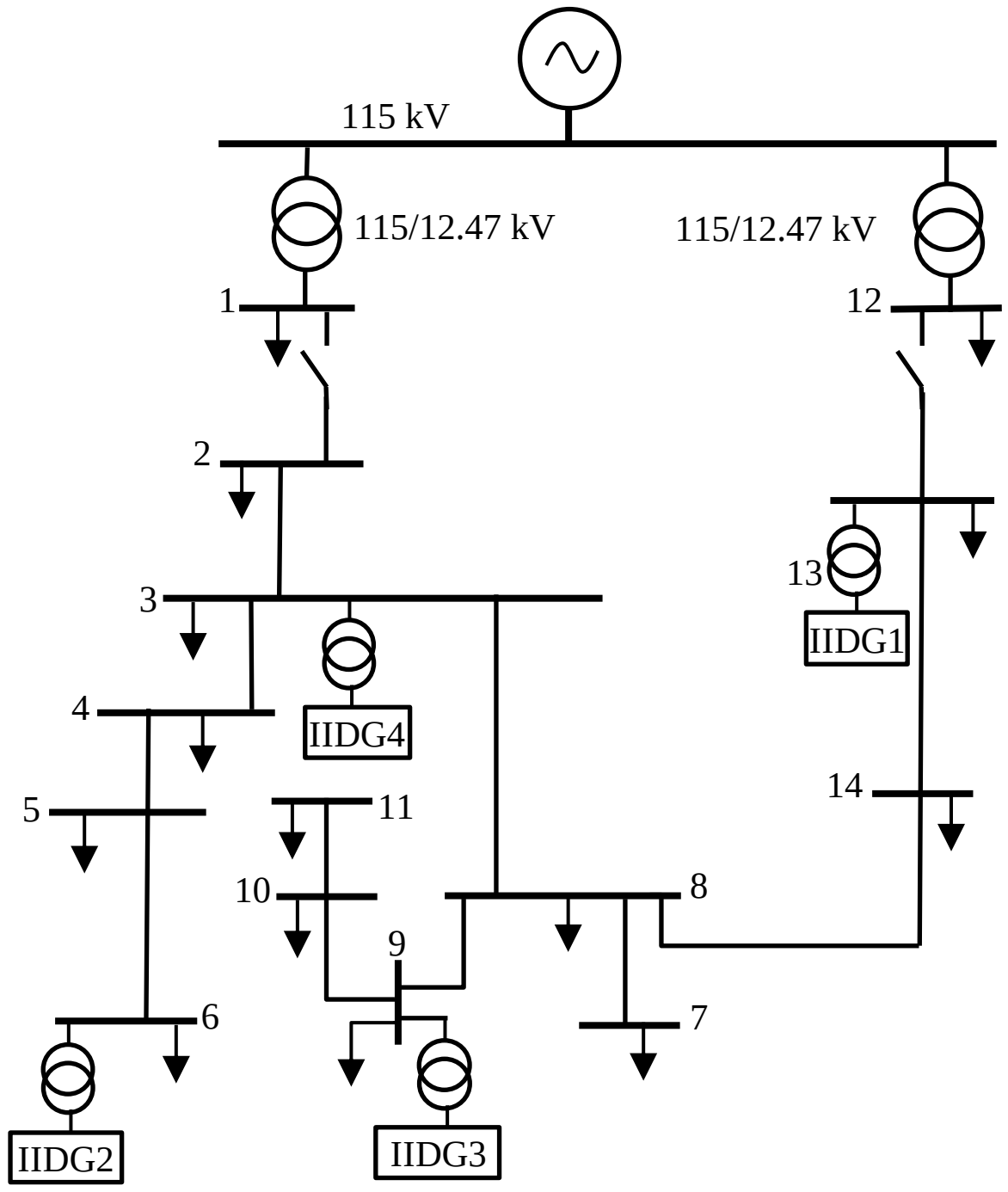

Fig.2. Schematic of the CIGRE Medium Voltage (MV) test system

**Scenario 1:** *The proposed distributed cooperative controller is not utilized.*

In this scenario only conventional droop-based controllers are active, and the proposed controller is not utilized. The droop gains of all IIDGs are equal to $7 \times 10^{-6}\ V/Var$ . In this scenario, the Microgrid is at the steady state condition before $t = 0.5\ sec$ when a load of 5 $MVA$ with power factor of 0.8 is added to Bus#4.

The voltage of the pilot bus (i.e., Bus#8) is shown in Fig. 3. As shown in Fig.3, due to the disturbance at $t = 0.5\ sec$, the voltage of the pilot bus deviates from the steady state value. As discussed before, the reason for this steady state error in the voltage value is the fact that droop controllers are proportional controllers and cannot eliminate the steady state error. According to Fig. 4, the IIDGs do not contribute to the voltage regulation proportional to their droop gains which means while the droop gains of all IIDGs are equal, their reactive power contributions (i.e., ΔQs) are not equal.

**Scenario 2:** *The proposed distributed controller is utilized and IIDGs exchange information with neighboring IIDGs through a ring communication architecture in the presence of a lossy communication link*

In this scenario the same event as Scenario 1 occurs. However, at $t = 0.55\ sec$ the proposed distributed cooperative controller is activated to regulate the voltage of the pilot bus. IIDGs exchange information with each other through a ring communication structure which means each IIDG only exchanges the information with its neighboring IIDGs.

The communication link between IIDG3 and IIDG2 is lossy in which the data packets are dropped every $t = 0.1\ sec$. This means the communication topology changes frequently as the communication link between IIDG3 and IIDG2 goes on and off every $t = 0.1\ sec$. Note that based on Theorem 2, the controller does not require the knowledge of dwell-time, in other words it works as the data packet drops happen with any frequency. As it is shown in Figs. 5-6, the proposed distributed cooperative controller successfully regulates the voltage of the pilot bus. As the droop gains of all IIDGs are equal, the reactive power contributions of IIDGs are equal as shown in Fig. 6.

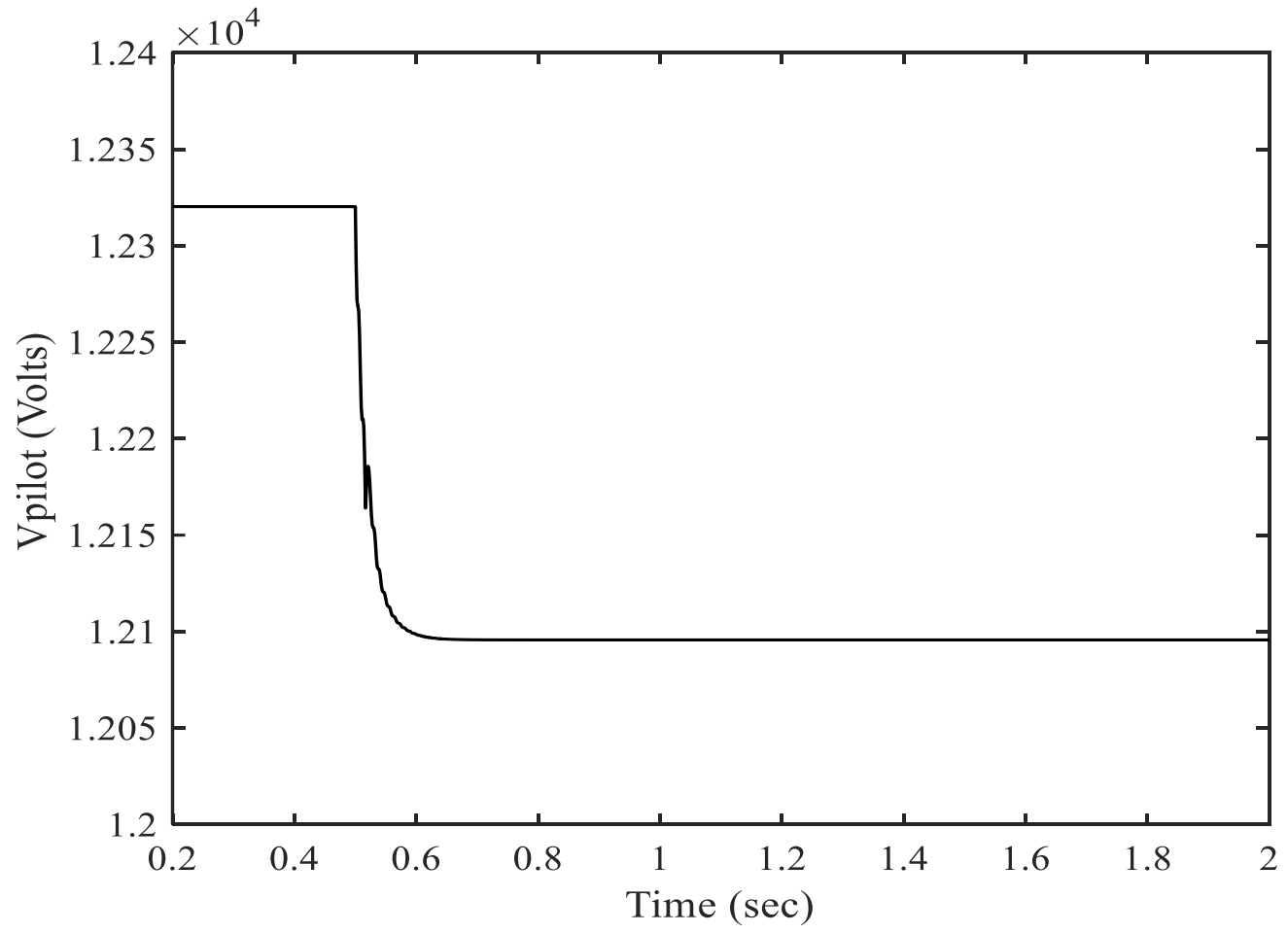

Fig.3. Voltage of the pilot bus in Scenario 1

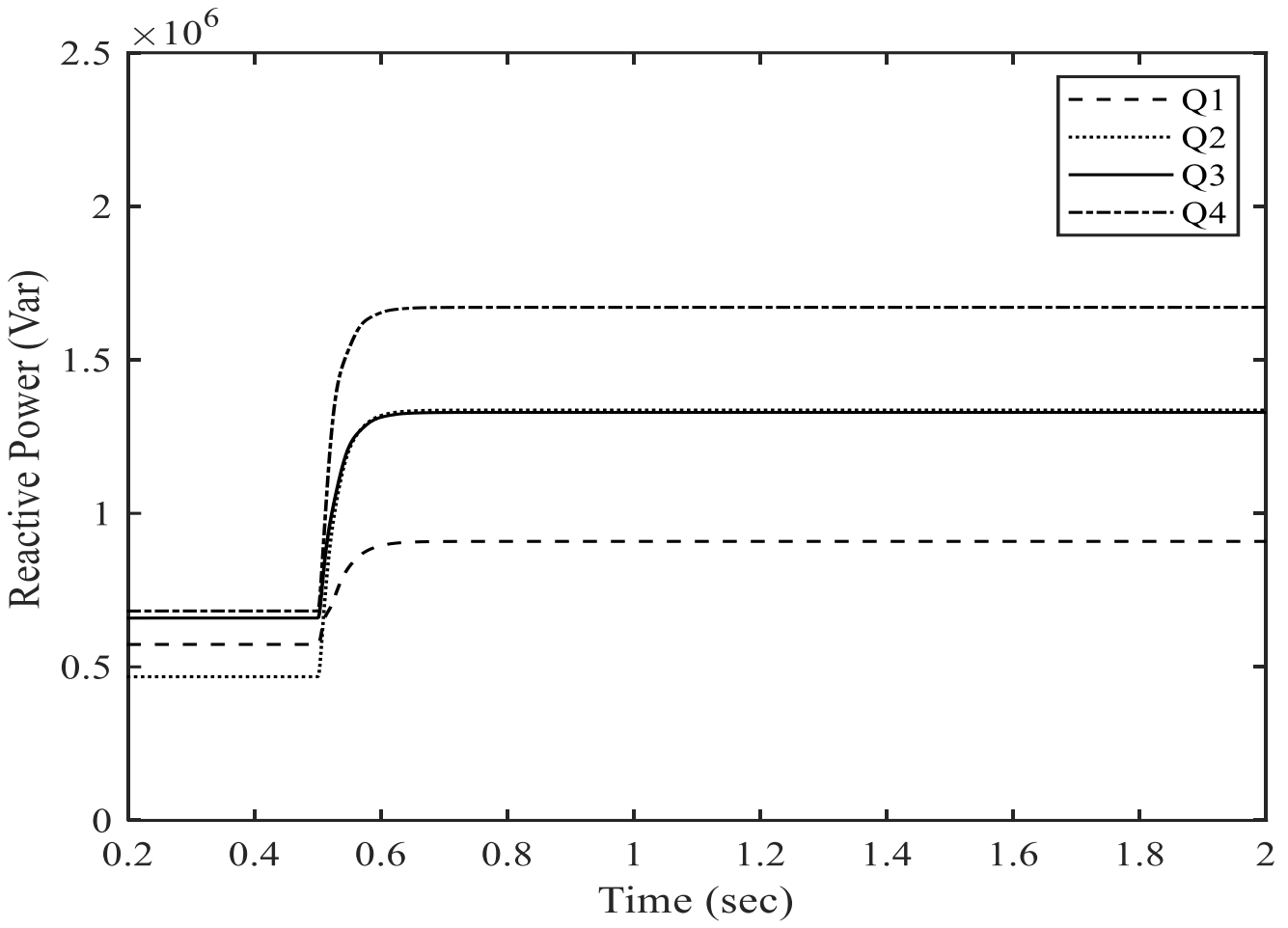

Fig. 4. Injected reactive power from IIDGs in Scenario 1

**Scenario 3:** *The communication link between IIDG2 and IIDG3 fails in the presence of multiplicative noise*

To demonstrate the performance of the proposed distributed cooperative voltage controller in facing the failure of the communication link, this scenario considers the outage of the communication link between IIGD2 and IIDG3 at $t = 0.8\ sec$. Utilizing the result of Theorem 1, multiplicative noise of $\pm 10\%$ in the communication network is considered. The condition of the system is the same as that of Scenario 2, but the droop gain value of IIDG4 is reduced to 80% that of IIGD1.

Therefore, IIDG4 should contribute more reactive power compared to other IIDGs. Fig.7 and 8 show the voltage of the pilot bus and the output reactive power of IIDGs. According to Figs.7-8 the proposed distributed cooperative voltage controller successfully regulates the voltage of the pilot bus even in the case of failure of the communication link.

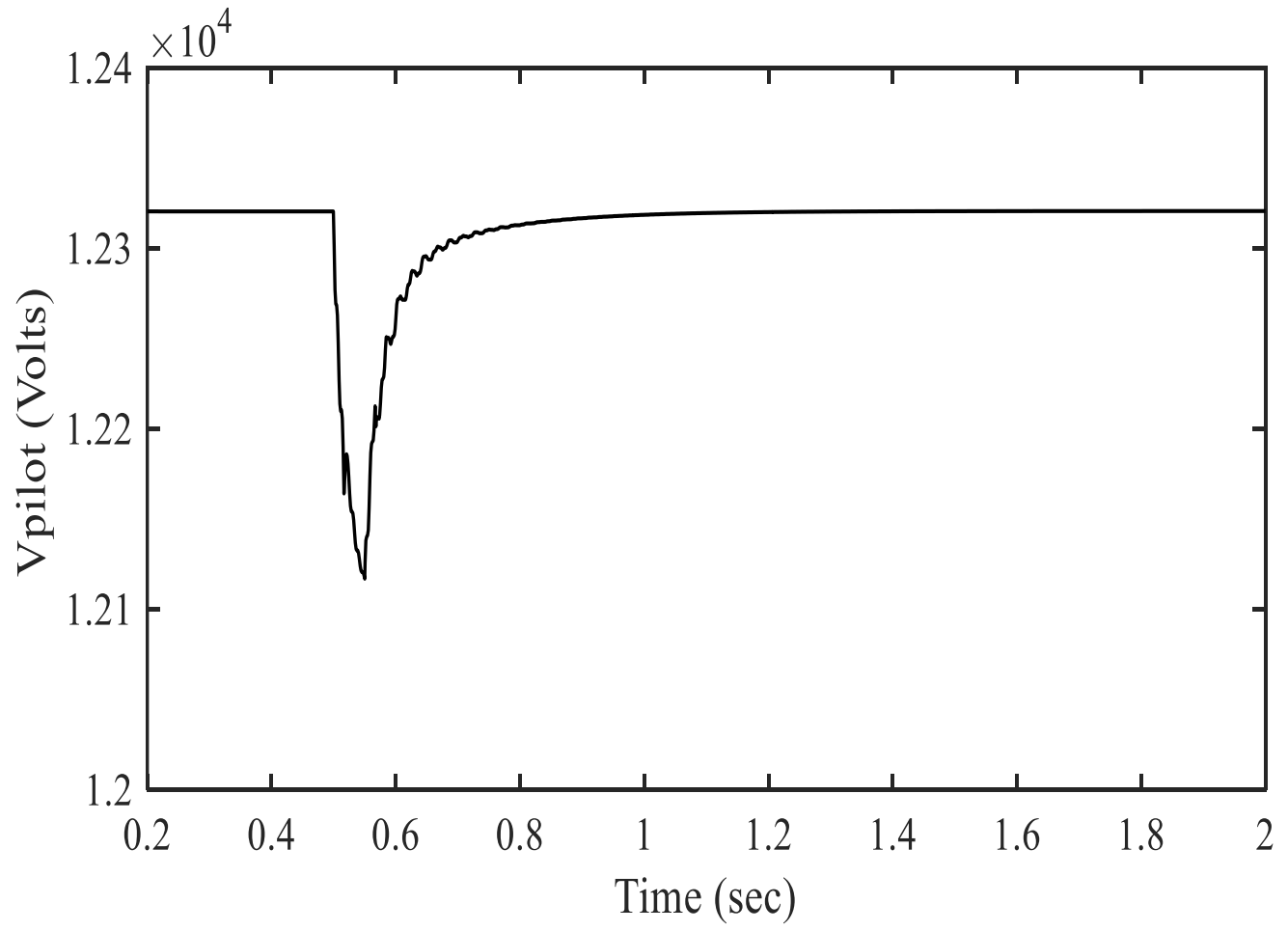

Fig.5. Voltage of the pilot bus in Scenario 2

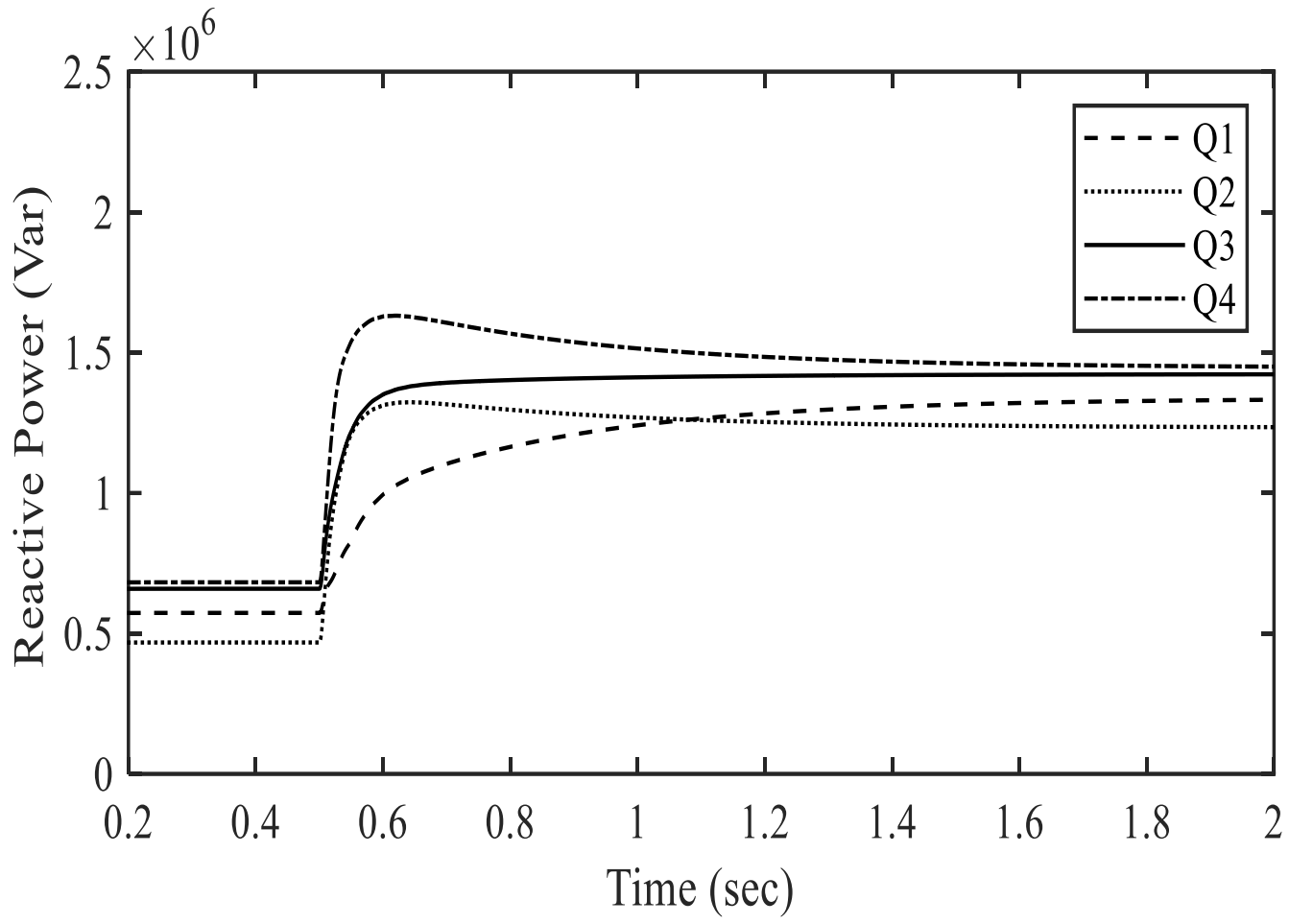

Fig.6. Injected reactive power from IIDGs in Scenario 2

As droop gains of IIDG1-IIDG3 are equal while the droop gain of IIDG4 is 80% of that of IIGD1, the reactive power contributions of IIDG1-IIDG3 are equal while the reactive power contribution of IIDG4 is more than IIDG1.

**Scenario 4:** *All the communication links between IIDG3 and the rest of the system fail and time delay of* $20\ msec$ *is considered in the communication links.*

This scenario demonstrates the performance of the proposed distributed cooperative voltage controller under a communication failure case that leads to the change in the number of IIDGs that are controlled by the distributed cooperative controller. The condition of the system is the same as that of Scenario 2. At $t = 0.6\ sec$ all communication links of IIDG3 fail. This may happen as the result of failure of transmitter/receiver systems at IIDG3. Once the communication link is lost, IIDG3 operates only based on the droop controller and latest received control command from the distributed cooperative controller at $t = 0.6\ sec$. Also, time delay of $20\ msec$ in the communication system is considered.

Figs. 9-10 show the proposed scale-free distributed cooperative controller regulates the voltage of the pilot bus successfully even in the case of the communication failure that leads to change in the number of IIDGs controlled by the distributed cooperative controller.

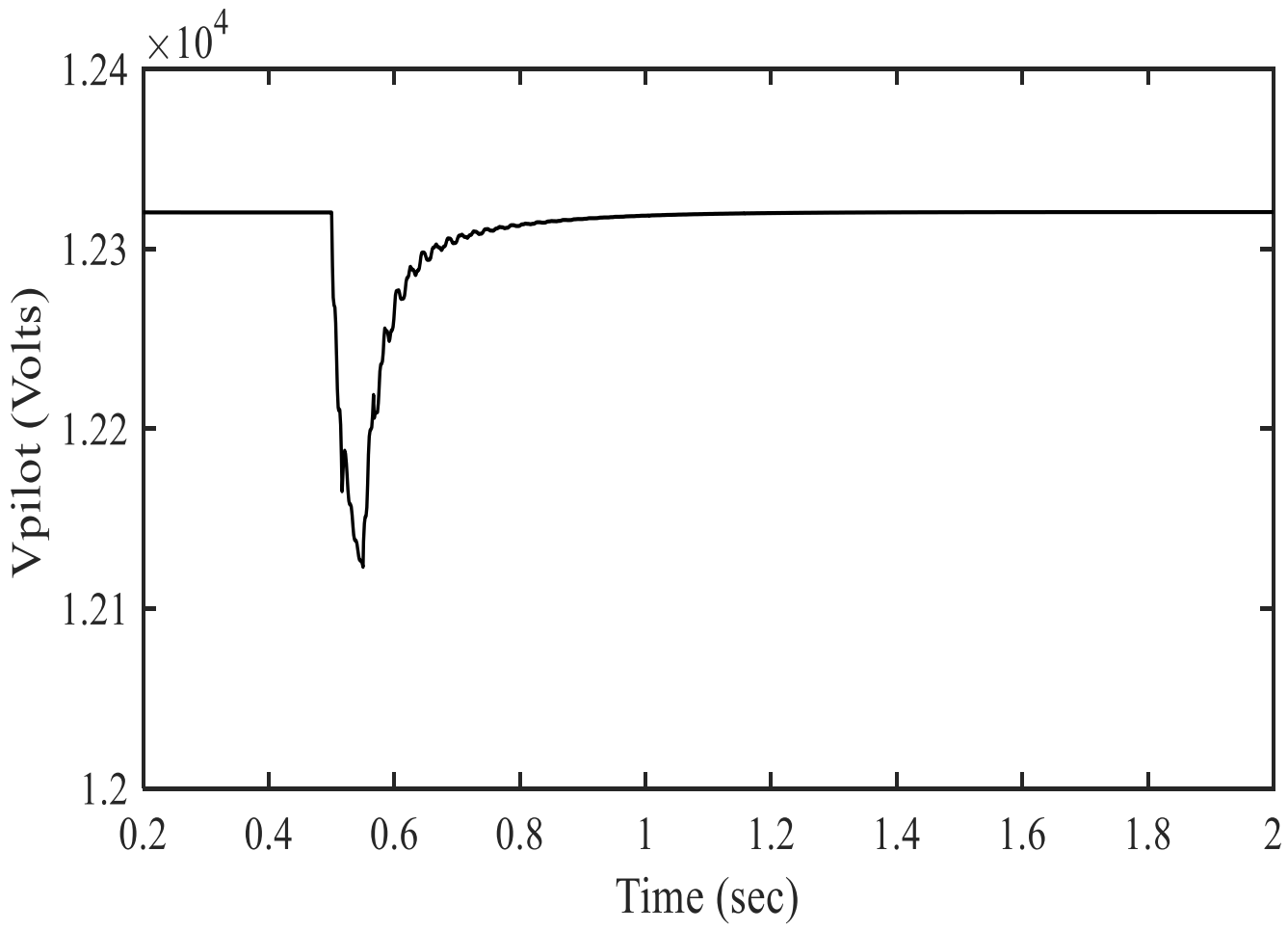

Fig. 7. Voltage of the pilot bus in Scenario 3

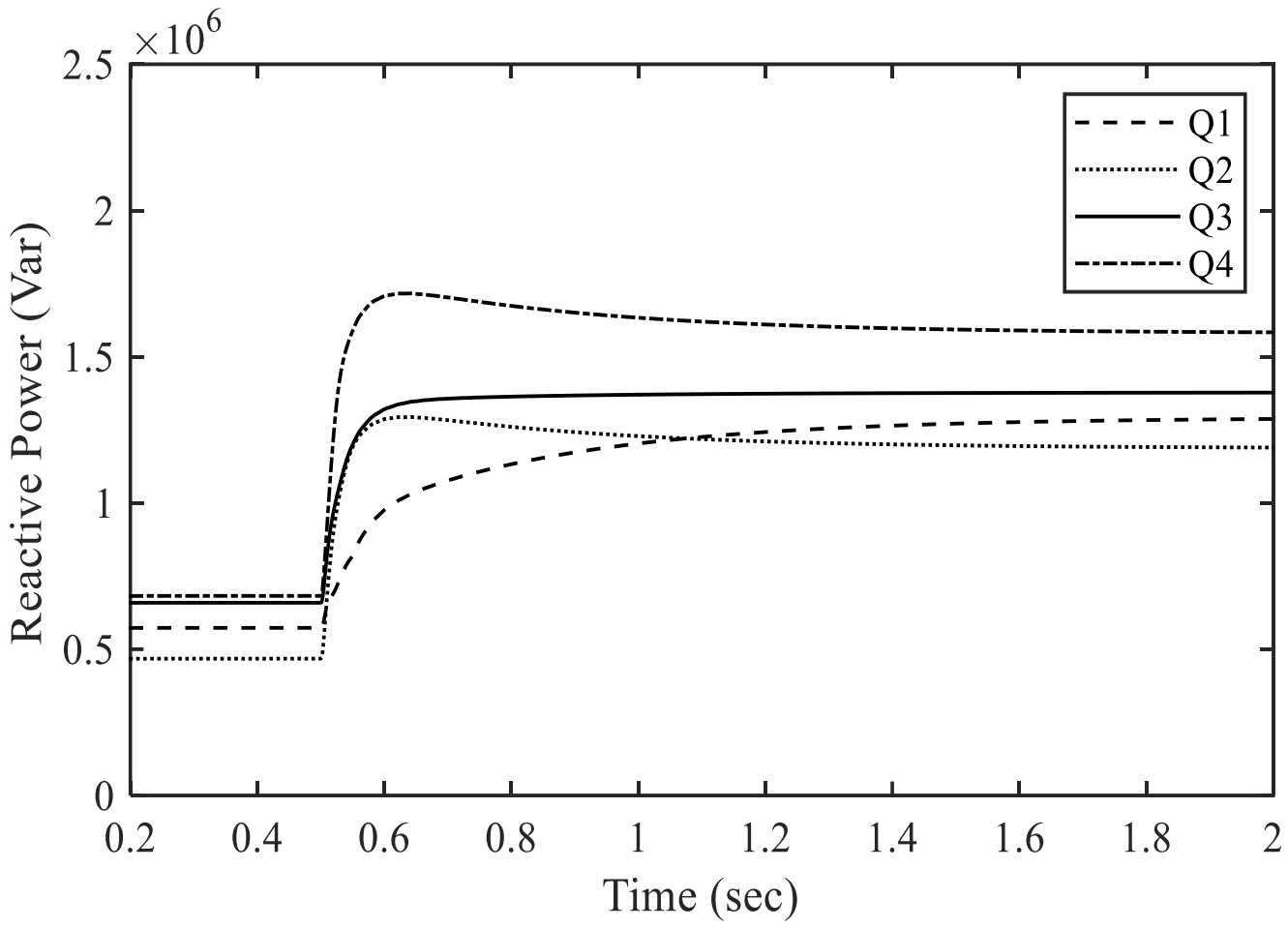

Fig.8. Injected reactive power from IIDGs in Scenario 3

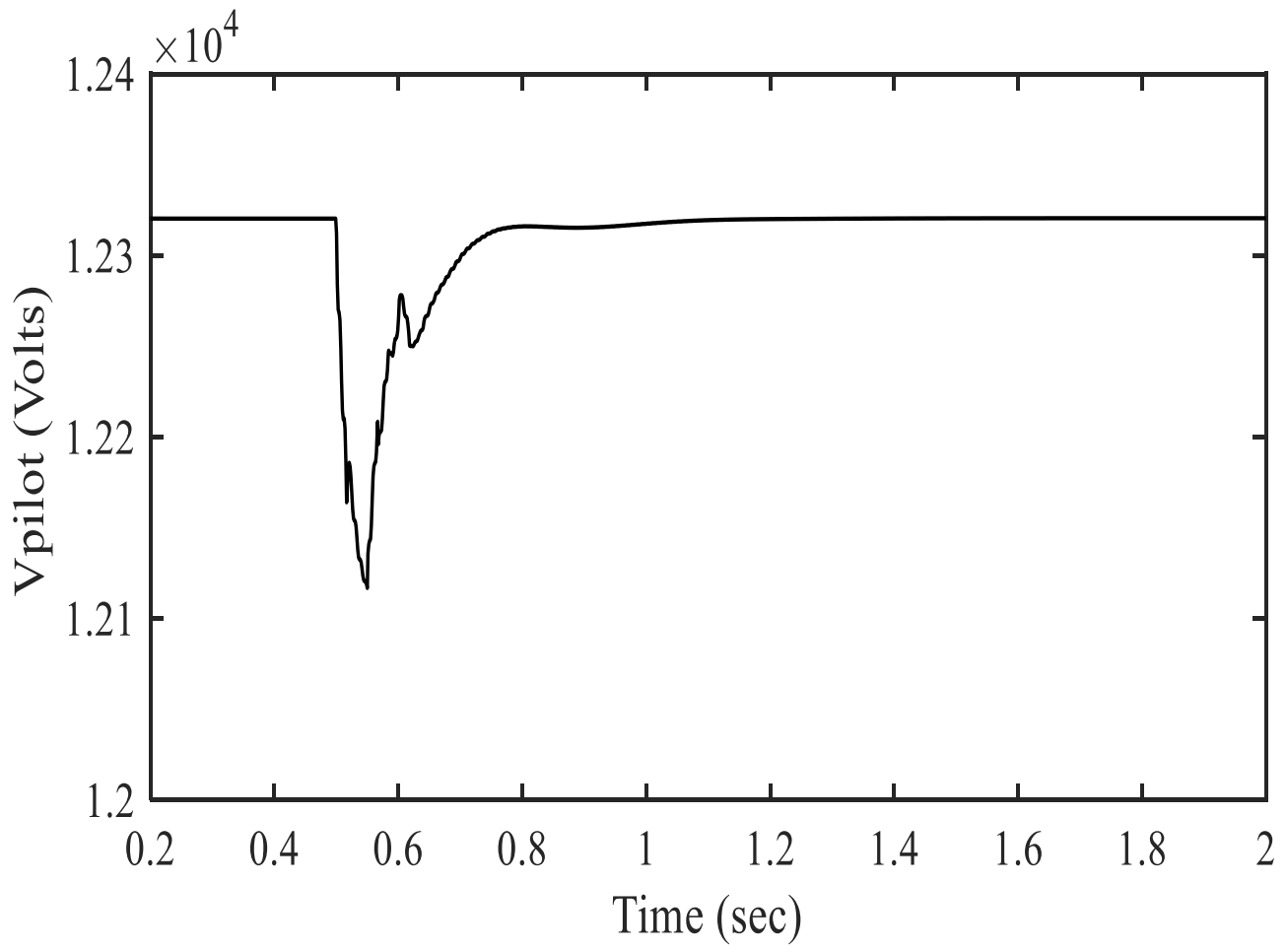

Fig.9. Voltage of the pilot bus in Scenario 4

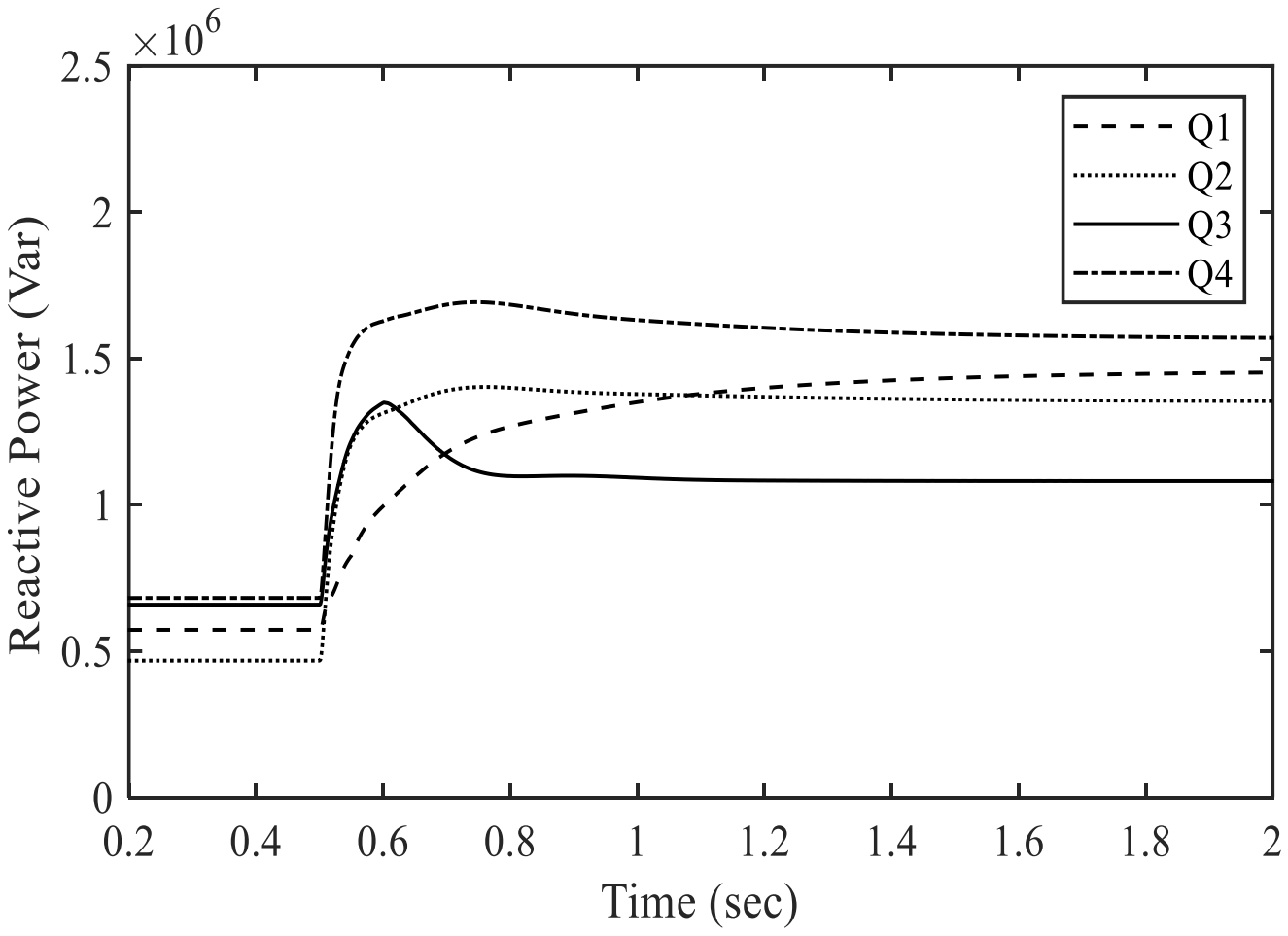

Fig.10. Injected reactive power from IIDGs in Scenario 4

Moreover, Fig. 10 shows that IIEDs1, IIEDs2, and IIEDs4 contribute to the voltage control task equally. This means they have contributed to the voltage control proportional to their droop gain values which are equal.

It should be noted, as expected, the contribution of IIDG3 is not synchronized with the rest of IIDGs as shown in Fig. 10. This is because, due to the communication failure, IIDG3 is not controller by the distributed cooperative controller anymore. Instead, it is controlled based on the droop control and the latest control command from the distributed cooperative controller before the communication failure. If the communication failure did happen, the results were the same as Fig. 6 in Scenario 2.

## V. Conclusion

In this paper a scale-free voltage controller for Microgrids based on distributed cooperative control was proposed. A double-integrator dynamic model for IIDGs was derived. The developed model was used in the proposed distributed cooperative control scheme to regulate the voltage of the pilot bus of the Microgrid. The communication topology of the Microgrid system was associated to a general time-varying graph which enhances the resilience of the proposed controller against communication link failures, data packet drops, and multiplicative noise.

Moreover, the stability of the system in the presence of unknown and arbitrarily communication delays was proved. Despite most of the available protocols in the literature the proposed controller is scale-free and independent from the information about the communication network such as bounds on the spectrum of the associated Laplacian matrix or the size of the network (i.e., the number of IIDGs). The performance of the proposed scale-free controller was studied by simulating the controller in the CIGRE medium voltage test system. The simulation results demonstrated that the proposed distributed cooperative voltage control can successfully regulate voltage of Microgrids even in the presence of communication failures, communication systems delays and lossy communication links.

## Appendix A

For the proof of Theorem 1, we need the following Lemma for the stability of time-delay systems.

**Lemma 1 [23, Lemma 3]** Consider a linear time-delay system

$$\dot{x} = Ax + \sum_{i=1}^{m} A_i x(t-\tau_i) \tag{A.1}$$

where $x(t) \in \mathbb{R}^n$ and $\tau_i \in \mathbb{R}_{\geq 0}$. Assume that $A + \sum_{i=1}^{m} A_i$ is Hurwitz stable. Then (A.1) is asymptotically stable for $\tau_1, \dots \tau_N \in [0, \bar{\tau}]$ if

$$det\left[j\omega I - A - \sum_{i=1}^{m} e^{j\omega\tau_i} A_i\right] \neq 0, \qquad \text{(A.2)}$$

for all $\omega \in \mathbb{R}$ and for all $\tau_1, \dots \tau_N \in [0, \bar{\tau}]$.

*Proof of Theorem 1:* We have

$$\dot{\bar{x}}^i = A\bar{x}^i + Bu^i$$
$$\bar{\zeta}^i = \sum_{j=1}^{N} \bar{\ell}_{ij}(t)\, \bar{x}^j(t - \tau_{ij}) \qquad \text{(A.3)}$$

with $\bar{x}^i = x^i - x^r$. Therefore, by combining (29) and (A.3), the closed loop system can be written as follows

$$\dot{\bar{x}}^i = A\bar{x}^i - \sum_{j=1}^{N-1} \bar{\ell}_{ij}(t)\rho^i BB^T P \bar{x}^j\left(t - \tau_{ij}\right). \qquad \text{(A.4)}$$

Then by defining

$$v^i = B^T P \bar{\zeta}^i$$

and

$$\bar{x} = \begin{pmatrix} \bar{x}^1 \\ \bar{x}^2 \\ \vdots \\ \bar{x}^N \end{pmatrix}, v = \begin{pmatrix} v^1 \\ v^2 \\ \vdots \\ v^N \end{pmatrix}, \rho = \begin{pmatrix} \rho^1 & \cdots & 0 \\ \vdots & \ddots & \vdots \\ 0 & \cdots & \rho^N \end{pmatrix}$$

and utilizing (29), it can be obtained in frequency domain

$$j\omega\bar{x} = \left[I \otimes A - \left(\rho(j\omega)\bar{L}_{j\omega}(\tau) \otimes (BB^T P)\right)\right]\bar{x} \qquad \text{(A.5)}$$

where $\bar{L}_{j\omega}(\tau) = L_{j\omega}(\tau) + diag\{\iota^i\}$ and

$$L_{j\omega}(\tau) = \begin{pmatrix} \ell_{11}(j\omega) & \ell_{21}(j\omega)e^{-j\omega\tau_{12}} & \cdots & \ell_{1N}(j\omega)e^{-j\omega\tau_{1N}} \\ \vdots & \ddots & \ddots & \vdots \\ \ell_{N1}(j\omega)e^{-j\omega\tau_{N1}} & \dots & \dots & \ell_{NN}(j\omega) \end{pmatrix}.$$

Following Lemma 1, the proof has two steps: 1) proving the stability of (A.5) without delay $\tau_{ij}$. 2) checking the determinant condition (A.2) for the stability of the system in presence of delays $\tau_{ij}$.

***Step 1:*** In the absence of delay, (A.5) would be equal to

$$\dot{\bar{x}} = [I \otimes A - (\rho\bar{L}(t)) \otimes (BB^T P)]\bar{x} \qquad \text{(A.6)}$$

By contradiction it implies that all $\rho^i$ are bounded when all $v^i \in \mathcal{L}_2$.

Assuming that the fact of all $v^i \in \mathcal{L}_2$ is not true, then there exist some agents with $\rho^i \to \infty$ as $t \to \infty$. Without loss of generality, it is assumed that $\rho^i \to \infty$ (i.e. $v^i \notin \mathcal{L}_2$) for agents $i = 1, \dots, k$, while for agents $i = k+1, \dots, N$, it can be obtained $\rho^i(t)$ is bounded (i.e. $v^i \in \mathcal{L}_2$). Then, define

$$\bar{x}^I = \begin{pmatrix} \bar{x}^1 \\ \bar{x}^2 \\ \vdots \\ \bar{x}^k \end{pmatrix}, v^I = \begin{pmatrix} v^1 \\ v^2 \\ \vdots \\ v^k \end{pmatrix}, \rho^I = \begin{pmatrix} \rho^1 & & & \\ & \rho^2 & & \\ & & \ddots & \\ & & & \rho^k \end{pmatrix},$$

$$\bar{x}^{II} = \begin{pmatrix} \bar{x}^{k+1} \\ \bar{x}^{k+1} \\ \vdots \\ \bar{x}^N \end{pmatrix}, v^{II} = \begin{pmatrix} v^{k+1} \\ v^{k+2} \\ \vdots \\ v^N \end{pmatrix}, \text{and}$$

$$\rho^{II} = \begin{pmatrix} \rho^{k+1} & & & \\ & \rho^{k+2} & & \\ & & \ddots & \\ & & & \rho^N \end{pmatrix}.$$

The following can be acquired

$$\bar{L}(t) = \begin{pmatrix} \bar{L}_1(t) & \bar{L}_2(t) \\ \bar{L}_2^T(t) & \bar{L}_3(t) \end{pmatrix}, \bar{L}^{-1}(t) = \begin{pmatrix} \bar{L}_a(t) & \bar{L}_b(t) \\ \bar{L}_b^T(t) & \bar{L}_c(t) \end{pmatrix}. \qquad \text{(A.7)}$$

Note that when $k = N$ i.e., all $\rho^i$ are unbounded, the above decomposition is not needed. The following proof is then still valid and all terms involving $x^{II}$, $v^{II}$ and $\rho^{II}$ are no longer present.

Next, the following candidate of Lyapunov function is used.

$$V^I = (\bar{x}^I)^T((\rho^I)^{-1} \otimes P)\bar{x}^I \qquad \text{(A.8)}$$

The time derivative of $V^I$ can be calculated as follows along the trajectories of system (A.3).

$$\dot{V}^I = -(\bar{x}^I)^T((\rho^I)^{-2}\dot{\rho}^I) \otimes P\bar{x}^I \qquad \text{(A.9)}$$

Note that

$$\bar{L}^{-1}(t)\begin{pmatrix} (\rho^I)^{-1} - 2\bar{L}_1(t) & -\bar{L}_2(t) \\ -\bar{L}_2^T(t) & 0 \end{pmatrix}\bar{L}^{-1}(t)$$
$$= \begin{pmatrix} \bar{L}_a(t)(\rho^I)^{-1}\bar{L}_a(t) - 2\bar{L}_a(t) & \bar{L}_a(t)(\rho^I)^{-1}\bar{L}_b(t) - \bar{L}_b(t) \\ \bar{L}_b^T(t)(\rho^I)^{-1}\bar{L}_a(t) - \bar{L}_b^T(t) & \bar{L}_b(t)(\rho^I)^{-1}\bar{L}_b(t) \end{pmatrix} \qquad \text{(A.10)}$$

Since $\bar{L}(t) \leq \gamma I$, it is known $\bar{L}^{-1}(t) \geq \gamma^{-1}I$ and also $\bar{L}_a(t)$, $\bar{L}_b(t)$, and $\bar{L}_c(t)$ are all bounded. Moreover, $\rho^I \to \infty$. As a result, there exists $T > 0$, and $\mu, \nu > 0$ such that

$$\begin{pmatrix} \bar{L}_a(t)(\rho^I)^{-1}\bar{L}_a(t) - 2\bar{L}_a(t) & \bar{L}_a(t)(\rho^I)^{-1}\bar{L}_b(t) - \bar{L}_b(t) \\ \bar{L}_b^T(t)(\rho^I)^{-1}\bar{L}_a(t) - \bar{L}_b^T(t) & \bar{L}_b(t)(\rho^I)^{-1}\bar{L}_b(t) \end{pmatrix} \leq \begin{pmatrix} -\mu I & 0 \\ 0 & \nu I \end{pmatrix} \qquad \text{(A.11)}$$

for $t > T$ which yields

$$\bar{L}^{-1}(t)\begin{pmatrix} (\rho^I)^{-1} - 2\bar{L}_1(t) & -\bar{L}_2(t) \\ -\bar{L}_2{}^T(t) & 0 \end{pmatrix}\bar{L}^{-1}(t) \leq \begin{pmatrix} -\mu I & 0 \\ 0 & \nu I \end{pmatrix}$$

Let $\alpha$ be such that $M > \alpha P$. Using this bound in (A.9) for the first term on the right and using (A.10) for the second term on the right, the following can be obtained

$$\dot{V}^I \leq -\alpha V - \mu(v^I)^T v^I + \nu(v^{II})^T v^{II} \qquad \text{(A.12)}$$

for $t > T$. By construction, it is obtained that $V^I \notin \mathcal{L}_2$ and $V^{II} \in \mathcal{L}_2$. But this yields a contradiction with inequality (A.12) since $V^I$ is always nonnegative. It implies that all the $v^i \in \mathcal{L}_2$ or equivalently, all the $\rho^i$ are bounded. Then, for the following Lyapunov function

$$V = \bar{x}^T(\rho^{-1} \otimes P)\bar{x} \qquad \text{(A.13)}$$

the time derivative can be calculated as follows

$$\dot{V} = -\bar{x}^T((\rho^{-2}\dot{\rho}) \otimes P)\bar{x} + \bar{x}^T(\rho^{-1} \otimes (PA + A^T P))\bar{x} - 2\bar{x}^T(\bar{L}(t) \otimes (PBB^T P))\bar{x} \qquad \text{(A.14)}$$

which yields, using similar techniques as before

$$\dot{V} = -\alpha V + m v^T v \qquad \text{(A.15)}$$

for some $m > 0$. Since $v \in \mathcal{L}_2$ it is obtained that $V(t) \to 0$ as $t \to \infty$. Meanwhile, since $\rho$ is bounded it implies that $\bar{x} \to 0$ as $t \to \infty$, i.e. $x^i \to x^r$. Namely, voltage regulation is achieved.

***Step 2:*** Next, in the light of Lemma 1, the closed-loop system (A.5) is asymptotically stable for all communication delays $\tau_{ij} \in \mathbb{R}_{\geq 0}$, if

$$det\left[j\omega I - \left[I \otimes A - \left(\rho(j\omega)\bar{L}_{j\omega}(\tau) \otimes (BB^T P)\right)\right]\right] \neq 0 \qquad \text{(A.16)}$$

for all $\omega \in \mathbb{R}$ and any communication delays $\tau_{ij} \in \mathbb{R}_{\geq 0}$.

Condition (A.16) is satisfied if matrix

$$\left[I \otimes A - \left(\rho(j\omega)\bar{L}_{j\omega}(\tau) \otimes (BB^T P)\right)\right] \quad \text{(A.17)}$$

has no eigenvalues on the imaginary axis for all $\omega \in \mathbb{R}$. That is to say we just need to prove the stability of (A.17). Since we have $\left|e^{-j\omega\tau_{ij}}\right| = 1$ and $\bar{L}(t)$ is a symmetric matrix, one can obtain the eigenvalues of $L_{j\omega}(\tau)$ have the positive real part. Correspondingly, we also have

$$L_{j\omega,1}(\tau) + L_{j\omega,1}^T(\tau) > 0 \quad \text{(A.18)}$$

Thus, similar to the proof of step 1, we can obtain the stability of (A.17). Then condition (A.16) is satisfied. Therefore, based on Lemma 1, for all $\tau_{ij}$

$$\bar{x}^i \to x^r,$$

as $t \to \infty$. In other words, voltage regulation is achieved.

## APPENDIX B

**Remark 2** *Let $\mathcal{G}_{\sigma(t)} \in \mathbb{G}_N$. Then from [24, Lemma 7] it follows that $\mathcal{G}_{\sigma(t)} \in \mathbb{G}_{\beta,\gamma,N}$ for some $\gamma > \beta > 0$. This leads to the Theorem 1.*

*Proof of Theorem 2:* The proof follows from the proof of Theorem 1 and Remark 2.

## REFERENCES

[1]. M. Azimi, and S. Lotfifard, “A Nonlinear Controller Design for Power Conversion Units in Islanded Micro-grids using Interconnection and Damping Assignment tracking control” *IEEE Transactions on Sustainable Energy*, in press.

[2]. M. Khederzadeh, and H. Maleki “Frequency Control of Microgrids in Autonomous Mode by a Novel Control Scheme Based on Droop Characteristics” *Electric Power Components and Systems*, vol. 41, no.1, pp. 16-30, 2013.

[3]. T. Zhao, Z. Li, and Z. Ding, “Consensus-Based Distributed Optimal Energy Management with Less Communication in a Microgrid” *IEEE Transactions on Industrial Informatics*, vol. 15, no. 6, pp.3356-3367, 2019.

[4]. A. Bidram, A. Davoudi, F. L. Lewis, and J M. Guerrero “Distributed cooperative secondary control of microgrids using feedback linearization.” *IEEE Transactions on Power Systems*, vol. 28, no. 3, pp. 3462-3470, 2013.

[5]. G. Loui, W. Gu, W. Sheng, X. Song, and F. Gao “Distributed model predictive secondary voltage control of islanded Microgrids with feedback linearization.” *IEEE Access*, vol. 6, pp. 50169-50178, 2018.

[6]. J. W. Simpson-Porco, Q. Shafiee, F. Dörfler, J. C. Vasquez, J. M. Guerrero, and F. Bullo “Secondary frequency and voltage control of islanded microgrids via distributed averaging." *IEEE Transactions on Industrial Electronics*, vol. 62, no. 11, pp. 7025-7038, 2015.

[7]. X. Lu, X. Yu, J. Lai, J. M. Guerrero, and Hong Zhou “Distributed secondary voltage and frequency control for islanded microgrids with uncertain communication links.” *IEEE Transactions on Industrial Informatics*, vol. 13, no. 2, pp.448-460, 2016.

[8]. J. Lai, H. Zhou, X. Lu, X. Yu, and W. Hu “Droop-based distributed cooperative control for microgrids with time-varying delays.” *IEEE Transactions on Smart Grid*, vol. 7, no. 4, pp. 1775-1789, 2016.

[9]. S. Lotfifard, D. Nojavanzadeh, Z. Liu, A. Saberi, and A. A. Stoorvogel, “Distributed Cooperative Voltage Control of Multi-terminal HVDC Systems.” *IEEE Systems Journal*, early access.

[10]. C. He, G. Hu, F. L. Lewis, and A. Davoudi “A distributed feedforward approach to cooperative control of AC microgrids.” *IEEE Transactions on Power Systems*, vol. 31, no. 5, pp. 4057-4067, 2015.

[11]. T. Morstyn, B. Hredzak, and V. G. Agelidis “Distributed cooperative control of microgrid storage.” *IEEE transactions on power systems*, vol. 30, no. 5, pp. 2780-2789, 2014.

[12]. M. A. Mahmud, M. J. Hossain, H. R. Pota, and A. M. T. Oo “Robust nonlinear distributed controller design for active and reactive power sharing in islanded microgrids.” *IEEE Transactions on Energy Conversion*, vol. 29, no. 4, pp. 893-903, 2014.

[13]. N. M. Dehkordi, N. Sadati, and M. Hamzeh “Fully distributed cooperative secondary frequency and voltage control of islanded microgrids.” *IEEE Transactions on Energy Conversion*, vol. 32, no. 2, pp. 675-685, 2016.

[14]. D. O. Amoateng, M. Al Hosani, M. S. Elmoursi, K. Turitsyn, and J. L. Kirtley “Adaptive voltage and frequency control of islanded multi-microgrids.” *IEEE Transactions on Power Systems*, vol. 33, no. 4, pp. 4454-4465, 2018.

[15]. J. H. Seo, J. Back, H. Kim, and H. Shim, “Output feedback consensus for high-order linear systems having uniform ranks under switching topology” *IET Control. Theory. Applications*, vol. 6, no. 8, pp.1118-1124, 2012.

[16]. N. Pogaku, M. Prodanovic, and T. C. Green, “Modeling, Analysis and Testing of Autonomous Operation of an Inverter-Based Microgrid” *IEEE Transactions on Power Electronics*, vol. .22, no. 2, pp. 613-625, 2007.

[17]. C. Godsil and G. Royle. Algebraic graph theory, volume 207 of Graduate Texts in Mathematics. Springer-Verlag, New York, 2001.

[18]. W. Ren and Y.C. Cao. Distributed coordination of multi-agent networks. Communications and Control Engineering. Springer-Verlag, London, 2011.

[19]. A. A. Stoorvogel, A. Saberi and M. Zhang “Synchronization in a homogeneous, time-varying network with nonuniform time-varying communication delays,” in Proc. 55th CDC, 2016, pp. 910–915.

[20]. D. Liberzon and A. S. Morse, “Basic problems in stability and design of switched systems,” *IEEE Contr. Syst. Magazine*, vol. 19, no. 5, pp. 59-70, 1999.

[21]. Benchmark Systems for Network Integration of Renewable and Distributed Energy Resources, CIGRE, Task Force C6.04, 2014.

[22]. R. Perez-Ibacache, C. A. Silva, and A. Yazdani, “Linear State-Feedback Primary Control for Enhanced Dynamic Response of AC Microgrids” *IEEE Transactions on Smart Grid*, vol. 10, no. 3, pp. 3149-3161, 2019.

[23]. M. Zhang, A. Saberi, and A. A. Stoorvogel “Synchronization in a network of identical continuous ‐ or discrete ‐ time agents with unknown nonuniform constant input delay” *International journal of robust and nonlinear control* vol. 28, no. 13 pp. 3959-3973, 2018.

[24]. H. F. Grip, T. Yang, A. Saberi and A. A. Stoorvogel “Output synchronization for heterogeneous networks of non-introspective agents” *Automatica*, vol. .48, no. 10, pp. 2444–2453, 2012.